\documentclass[aps,prb,twocolumn,superscriptaddress]{revtex4-2}
\usepackage{graphics,graphicx, array, afterpage}
\usepackage{qcircuit,amsmath}
\usepackage{grffile}
\usepackage[export]{adjustbox}
\usepackage{lineno}
\usepackage{xcolor}
\usepackage{longtable}
\usepackage{booktabs}
\usepackage{braket}
\usepackage{array}
\usepackage{multirow}
\usepackage{enumerate}
\usepackage{amsmath}
\usepackage{amssymb}
\usepackage{amsthm}
\usepackage{times}
\usepackage{multirow}

\definecolor{LinkColor}{RGB}{0,0,128} %{192,77,77}
\usepackage[colorlinks, citecolor=blue]{hyperref}
\hypersetup{linkcolor=LinkColor,urlcolor=LinkColor,citecolor=LinkColor,pdfstartview={FitH},urlcolor=LinkColor}

\usepackage[utf8]{inputenc}
\usepackage{color}
\usepackage{tabularx}
\usepackage{algorithm}
\usepackage{algpseudocode}
\usepackage{graphbox}
\usepackage{amsfonts}
\usepackage{dcolumn}
\usepackage{bm}
\usepackage{epstopdf}

\begin{document}
\preprint{APS/123-QED}

\title{Entanglement phases and phase transitions in the monitored free fermion systems of localization}

\author{Yu-Jun Zhao}
\affiliation{School of Physics and Optoelectronics, Xiangtan University, Xiangtan 411105, China}
\affiliation{Institute for Quantum Science and Technology, Shanghai University, Shanghai 200444, China}

\author{Xuyang Huang}
\affiliation{Institute for Quantum Science and Technology, Shanghai University, Shanghai 200444, China}

\author{Yi-Rui Zhang}
\affiliation{Institute for Quantum Science and Technology, Shanghai University, Shanghai 200444, China}

\author{Han-Ze Li}
\email{hanzeli@u.nus.edu}
\affiliation{Institute for Quantum Science and Technology, Shanghai University, Shanghai 200444, China}
\affiliation{Department of Physics, National University of Singapore, Singapore 117542}

\author{Jian-Xin Zhong}
\email{jxzhong@shu.edu.cn}
\affiliation{Institute for Quantum Science and Technology, Shanghai University, Shanghai 200444, China}
\affiliation{School of Physics and Optoelectronics, Xiangtan University, Xiangtan 411105, China}

\begin{abstract}
The interplay between quantum measurement and localization significantly alters the spreading of quantum information. In this work, we investigate monitored free-fermion chains with localized potentials. With the aid of the quantum trajectory method and finite size analysis, we numerically reveal that the Berezinskii–Kosterlitz–Thouless transition, which emerges in monitored free-fermion chains, is robust in the presence of localization. To understand this, we construct a phase diagram that diverges at a boundary described by entanglement propagation. We find that the monitored system wtih Stark-localized decays quickly under small measurement strength, whereas the Anderson-localized system decays more slowly. 
% The presence of localized potentials has significantly enriched and diversified the entanglement patterns in monitored free fermion systems. In this work, we employ the stochastic Schr\"odinger equation to simulate a one-dimensional spinless fermion system under continuous measurement and localized potentials. By averaging the steady-state entanglement entropy over many quantum trajectories, we investigate its dependence on measurement and localization parameters. We used a phenomenological model to interpret the numerical results,  show that Stark potentials do not destroy the Berezinskii–Kosterlitz–Thouless universality class of the entanglement phase transition. The phase boundary is jointly determined by measurement and localized  potentials, depending on the induced localization mechanism. Distinct mechanisms yield different boundaries, while identical mechanisms characterize the boundary in the same way.
By incorporating localization effects, this work advances our understanding of how measurement competes with the coherent spreading of quantum information. Our findings can potentially be realized in cold atom systems, trapped ions, and quantum dot arrays.

\end{abstract}

\maketitle
% \tableofcontents
\section{Introduction}
Classical paradigms of nonequilibrium quantum physics-such as thermalization under the eigenstate thermalization hypothesis~\cite{Deutsch1991,Srednicki1994,Rigol2008,Dalessio2016,Kim2014} and its counterpoint in the quantum Zeno effect~\cite{Misra1977,Itano1990,Facchi2002,Kofman2000,Facchi2001}, offer a striking tension: while coherent dynamics generically scramble information and build volume-law entanglement, frequent observations can freeze evolution by continually projecting local degrees of freedom. This tension naturally leads to the notion of a measurement-driven reorganization of quantum states: as the rate or strength of monitoring is increased, the balance between entanglement production and removal can qualitatively change, suggesting a sharp transition in the long-time entanglement structure. This phenomenon is now understood as a measurement-induced phase transition (MIPT)~\cite{Fisher_2023, Li_2025,ls1,ls2,ls3,ls4,ls5,Poboiko:2023yid}, in which entanglement scaling morphs from volume law to area law due to the backaction of measurements.

A broad body of work has established both the ubiquity and diversity of MIPTs across platforms and protocols~\cite{Fisher2023ARCMP,Han2022PRB,Turkeshi2021PRB,Sierant2022Quantum,Sierant2022PRL,Sharma2022SciPostCore,Turkeshi2022PRB,Sierant2022PRB,Paviglianiti2024Quantum,Ladewig2022PRResearch,Buchhold2022arXiv,Botzung2023arXiv,Wang2024arXiv,Xiao2025arXiv,Jin2022PRL,Jin2024PRB,Lee2024PRXQ,Adani2024SciRep,Li2018,Chan2019,Skinner2019,Li2019,Gullans2020a,Gullans2020b,Jian2020,Bao2020,Choi2020,Szyniszewski2019,Fan2021,Vijay2020,Lavasani2021,Sang2021,Ippoliti2021,Tang2020,Goto2020,Fuji2020,Rossini2020,Lunt2020,Szyniszewski2023PRBB54309,3z,Agrawal2022PRX,Barratt2022PRL,Igo_prx,Lang2020,Coppola2022,Szyniszewski2020,DeTomasi2024,Yang2023}. In random quantum circuits~\cite{Li2018,Chan2019,Skinner2019,Li2019,Gullans2020a,Gullans2020b,Jian2020,Bao2020,Choi2020,Szyniszewski2019,Fan2021,Vijay2020,Lavasani2021,Sang2021,Ippoliti2021}, interleaving unitary gates with stochastic measurements yields a finite-threshold transition. Related transitions appear under Hamiltonian dynamics with projective or continuous   monitoring~\cite{Tang2020,Goto2020,Fuji2020,Rossini2020,Lunt2020}, and persist for both projective and indirect
measurements. Across these settings, the steady-state phases (volume, logarithmic/sub-volume, and area) and their critical properties depend on symmetries~\cite{Lavasani2021,Agrawal2022PRX,Barratt2022PRL,Han2022PRB},  dimensionality ~\cite{Igo_prx}, and the microscopic structure of measurements~\cite{Chan2019,Lang2020,Coppola2022,Szyniszewski2020,DeTomasi2024,Yang2023}, revealing a rich taxonomy of measurement-driven universality classes.
\begin{figure}[bt]
\hspace*{-0.45\textwidth}
\includegraphics[width=0.45\textwidth]{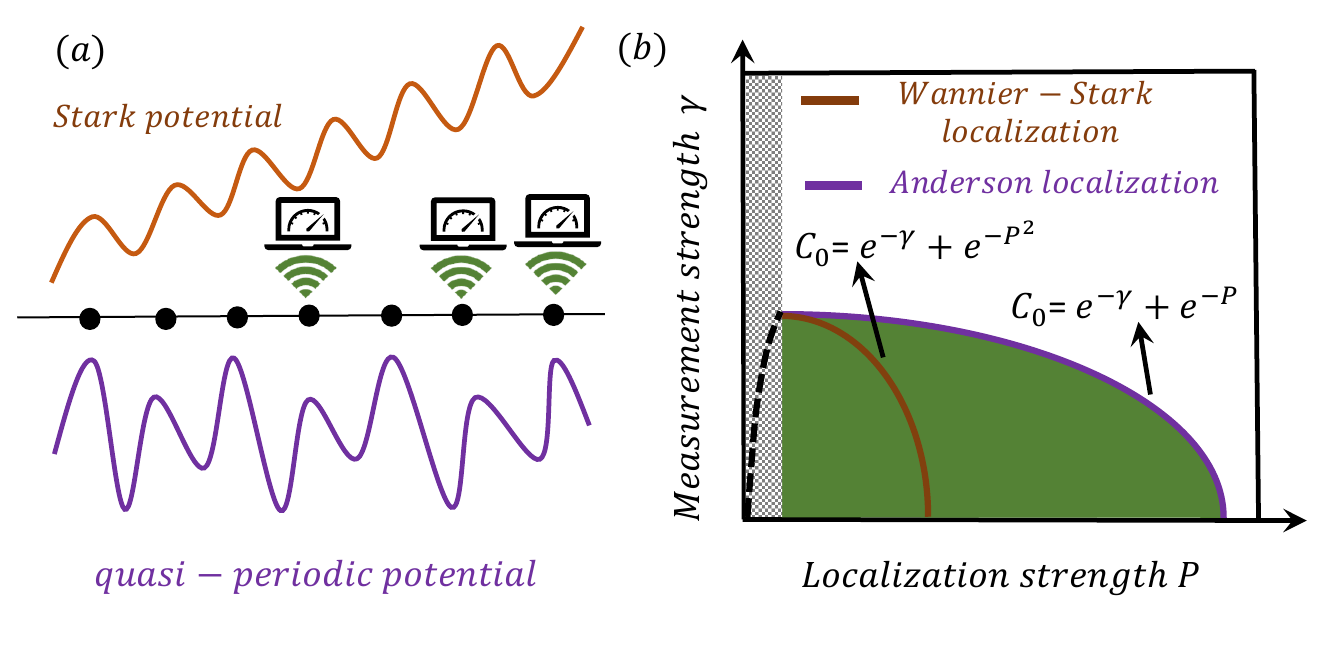} 
\caption{(a): Monitored half-filled fermions in SP (brown) and QPP (purple). (b): The phase boundaries in the phase diagram are formed by different localization mechanisms (Wannier-Stark localization and Anderson localization). We present the general solution form, with specific parameters given below. The grey region corresponds to the case of a very weak localization potential, where we tend to revert to the results obtained from the Keldysh field theory~\cite{Igo_prx}; the black dashed line corresponds to the phase boundary within this region.}\label{model}
\end{figure}

In the presence of localization, the landscape becomes more intricate. The root cause lies in the fact that localization tends to confine particles or quantum information to local regions, thereby significantly suppressing the generation and propagation of long-range entanglement. Different types of localization affect entanglement and information spreading through their respective physical mechanisms, and may thus alter the stability of phases and critical behavior in different ways. For example, disorder or quasi-periodic potential (QPP) induced Anderson localization and Stark potential (SP) induced Stark localization~\cite{PhysRevA.108.043301,PhysRevLett.130.120403,PhysRevB.111.014315} imprint distinct single-particle spectra, spatial profiles and transport properties that can~\cite{Anderson1958,Thouless1972,Abrahams1979,Wannier_1960_PhysRev,Gluck_2002_PhysRep, Zhong1995Quantum,Zhong2001Shape,Maksimov_2015_PhysRevA,Longhi_2009_PRL,Burkle_2001_PhysRevB,Hartmann_2004_NJP,Li_2024_PhysRevB,Zhao_2025_PhysRevB,Huang_2025_PhysRevB,Lye2007PRA,Roati2008Nature,DErrico2014PRL,Schreiber2015Science,Bordia2017PRX,An2021PRL,Harper1955PPSA,Aubry1980AIPS}, in principle, reshape measurement-driven entanglement phases and transitions. 

Previous studies in disordered~\cite{Szyniszewski2023PRBB54309} and quasi-periodic systems~\cite{3z} have revealed the characteristics of measurement-induced phase transitions, with their results mainly emphasizing the persistence  of the BKT universality class and the regulation of phase laws by Anderson localization potentials. In contrast to these studies based on Anderson localization, we choose the Stark localization system as a stronger localization platform, taking this as an opportunity to further explore the competition between measurement and different localization mechanisms, and attempt to construct a phase diagram that diverges at a boundary characterized by entanglement propagation under distinct localization mechanisms.

Against this backdrop, we ask two connected questions at the intersection of measurement and localization: (I) Will a change in the localization mechanism alter the critical universality class of the MIPT phase diagram, or do the two cases follow the same critical laws?  (II) Is it possible to establish a unified phase diagram to characterize entanglement phase transitions arising from the interplay between different localization mechanisms and measurements?  Addressing these questions disentangles universal from non-universal ingredients in the measurement-localization competition and yields operational criteria for experimental tests across platforms ranging from cold atomic systems~\cite{Goto2020,Corcovilos2019,Diehl2008,Tang2020,Fuji2020,Alberton2021} and trapped ions~\cite{Noel2022,Agrawal2024} to quantum dot arrays~\cite{Kim2022}.

To address question (I), we employ a one-dimensional free-fermion system as our platform. Under both a SP (realizing Stark localization) and a QPP (realizing Anderson localization), we introduce random projective measurements with tunable strength. By analyzing the system's steady-state entanglement entropy and effective central charge, we investigate the impact of the measurement-localization coupling on the entanglement phase transition. For question (II), we identify the physical mechanisms that play a central role in the entanglement phase transition, thereby formulating a universal statement concerning different localization mechanisms. Furthermore, we obtain several physical pictures, which are summarized schematically in Fig.~\ref{model}.

The remainder of this paper is organized as follows. In Sec.~\ref{sec2}, we introduce the tight-binding model in the presence of a localization potential and outline the quantum trajectory method used to simulate continuous measurement dynamics. In Sec.~\ref{sec3}, we present our understanding of the phase boundary arising from the interplay between different localization mechanisms and measurements, together with the entanglement phase diagram of MIPTs and the semi-analytical phase boundary we obtain. We then investigate the central charge, connected correlation function, and critical universality classes under both SP and QPP. Finally, Sec.~\ref{sec4} summarizes the entire work.
\section{model and method}\label{sec2}
\begin{figure}[bt]
\hspace*{-0.52\textwidth}
\includegraphics[width=0.5\textwidth]{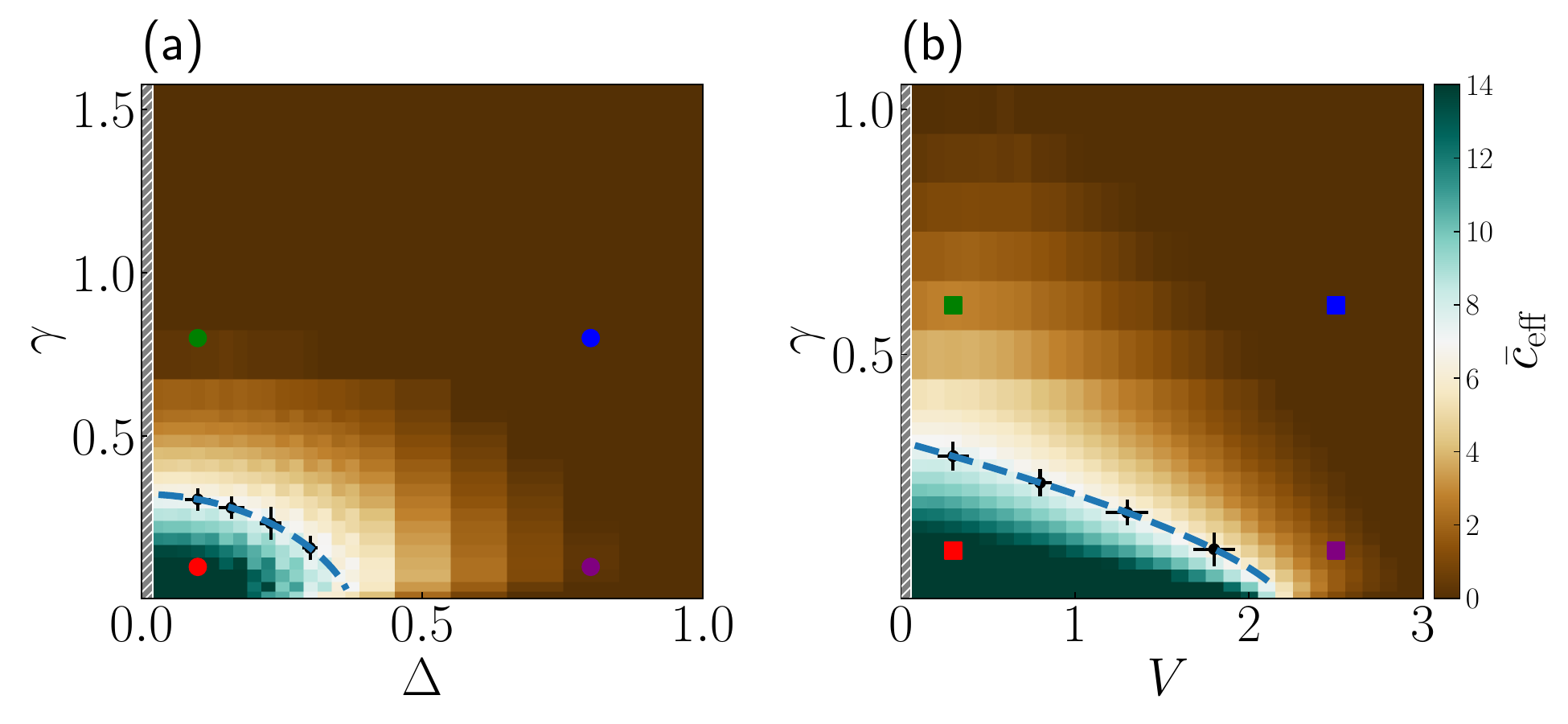} 
\caption{Phase diagrams and phase boundaries extracted from the effective central charge.  
(a) Phase diagram under joint tuning of the measurement strength and SP strength;  
(b) Phase diagram under joint tuning of the measurement strength and the QPP strength.  
Colors indicate the effective central charge \(\bar{c}_{\mathrm{eff}}\), extracted from the finite-size scaling of half-chain entanglement entropy: green corresponds to large \(\bar{c}_{\mathrm{eff}}\) values (log-law-like regime), while brown corresponds to \(\bar{c}_{\mathrm{eff}} \rightarrow 0\) (area-law regime).  
Dashed curves denote the semi-analytical expressions for phase boundaries derived by Eq.~\eqref{fs}.  
Black solid dots mark four phase transition points obtained through finite-size scaling analysis and data collapse. The grey region corresponds to the case of a very weak localization potential, where we tend to revert to the results obtained from the Keldysh field theory~\cite{Igo_prx}; therefore, it is not shown in the diagram.  
In the subsequent finite-size scaling analysis, four representative points on each phase diagram (marked in red, green, blue, and purple, respectively) are selected to characterize the entanglement scaling behavior at the phase boundary of that point.
}\label{phase}
\end{figure}
We consider a spinless one-dimensional fermions model of localization, subject to continuous measurements:
\begin{equation}
H=-J\sum_j \left(c_j^\dagger c_{j+1}+\mathrm{h.c.}\right)
+\sum_j P_j\,n_j, \label{hami}
\end{equation}
where $P_j$ represents the strength of potential at lattice site $j$:
\begin{equation}
P_j= \begin{cases}
\Delta j/L, & \text{SP}, \\
V \cos \left( \dfrac{2 \pi j}{\tau}+\theta \right), & \text{QPP},\\
\varepsilon_j, & \text{Anderson disorder},
\end{cases}
\end{equation}
with $J=1$ is the strength of the nearest-neighbor hopping term, $c_j $ and $c^\dagger_j$ are the annihilation and creation operators of the spinless fermion at site $j$. Here, $\Delta$ denotes the SP strength, $L$ is the number of lattice sites in the system, and the parameters $V$,  $\tau$, and $\theta$ represent, respectively, the QPP strength, the golden mean $(\sqrt{5} + 1)/2$, and the phase shift of the potential. The on-site potential $\varepsilon_j$ is an independent random variable drawn from the uniform distribution $\left[ -{W}, {W} \right]$, where $W$ is the disorder strength. We initialize the system in a N\'eel state, that is, $\psi(t=0)=$ $\prod_{i=1}^{L / 2} c_{2 i-1}^{\dagger}|\mathrm{vac}\rangle$, where $|\mathrm{vac}\rangle$ is the vacuum state, and consider its evolution under open boundary conditions (OBCs).We perform continuous measurements uniformly on all lattice sites with a finite measurement rate $\gamma$, where quantum-jump  events occur according to the standard quantum-jump trajectory formalism\cite{daley2014quantum, dalibard1992wave, wiseman1993interpretation, yamamoto2025measurement}
. 
The corresponding stochastic Schrödinger equation is given by~\cite{EA,Alberton2021,Fuji2020,wiseman2009quantum}
\begin{align}
d|\psi(t)\rangle &= -i H |\psi(t)\rangle \, dt \nonumber\\
&\quad + \sum_{j=1}^L \left[ 
\frac{c_j^{\dagger} c_j |\psi(t)\rangle}{\sqrt{\langle \psi(t)| n_j |\psi(t)\rangle}} 
- |\psi(t)\rangle \right] dW_j(t),
\end{align}\label{eq3}
where $n_j = c_j^{\dagger} c_j$ is the local particle number operator being monitored. 
The stochastic variables $dW_j(t)$ take values $0$ or $1$ according to independent Poisson processes with mean 
$\langle\!\langle dW_j(t)\rangle\!\rangle = \gamma \langle n_j\rangle dt,$
and $\langle\!\langle \cdots \rangle\!\rangle$ denotes the noise average. 
In this formulation, the Born probabilities are encoded in the normalization factor 
$\sqrt{\langle \psi(t)| n_j |\psi(t)\rangle}$, which ensures that the probability of each measurement outcome is consistent with quantum mechanics, while the Poisson increments $dW_j(t)$ determine whether a measurement event occurs at site $j$ during each infinitesimal time step. 
As a result, the system evolves along many possible quantum trajectories starting from the same initial state, and statistical properties of observables are obtained by averaging over a large ensemble of such trajectories without any postselect.  We use 500 independent trajectories in our numerical simulations.
The main physical quantity that we address is entanglement entropy. Divide the system into two subsystems $A$ and $B$, which have sizes $\ell$ and $L-\ell(\ell \leq L / 2)$, respectively ($L$ represents the system size). Then, the entanglement entropy is defined as:
\begin{align}
    S=-\operatorname{Tr}\left(\rho_A \log \rho_A\right),
\end{align}
where $\rho_A$ is the reduced density matrix of the subsystem $A$ for a given wave function $|\psi\rangle$, that is, $\rho_A=\operatorname{Tr}_B(|\psi\rangle\langle\psi|)$, where $\operatorname{Tr}_B$ is the partial trace with respect to the part $B$. $S$ is initially $0$ and increases over time, reaching a saturated saturated value in the long-time limit. We calculate the long-time steady-state entanglement entropy for each trajectory, take their average, and finally obtain $\bar{S}$.

% During the transition from volume law to area law, there are linear lines that indicate the entanglement entropy of the system around $L/2$ fits the functional form of conformal field theory (CFT), with an effective central charge $\bar{c}_{\rm eff}$:
% \begin{align}
%     \bar{S}(l, L)=\frac{c}{3} \ln \left(\frac{L}{\pi} \sin \frac{\pi l}{L}\right)+s_0,\label{eq4}
% \end{align}
% where $l$ is the length of the subsystem , $\bar{c}_{\rm eff}$ is the effective central charge of the non-unitary CFT, and $s_0$ is the residual entropy. For large enough systems, $\bar{c}_{\rm eff}$ is expected to be zero in the area law phase and finite in the log phase, and thus can be used as a transition diagnostic. However, even in free fermion circuits, where we can access larger system sizes, finite size effects are significant and impede our analysis. Special care needs to be taken for the critical phase, where both $\bar{S}$ and the correlation length $\xi$ diverge logarithmically with $L$--extraction of the critical point is difficult for phase transitions with slowly diverging length scales [33]. This critical phase is expected to be described by a 1+1D non-unitary conformal field theory (CFT) with OBC , as Eq.~\ref{eq4}.
\section{results}\label{sec3}
\subsection{Phase boundaries }
\begin{figure}[bt]
\hspace*{-0.53\textwidth}
\includegraphics[width=0.5\textwidth]{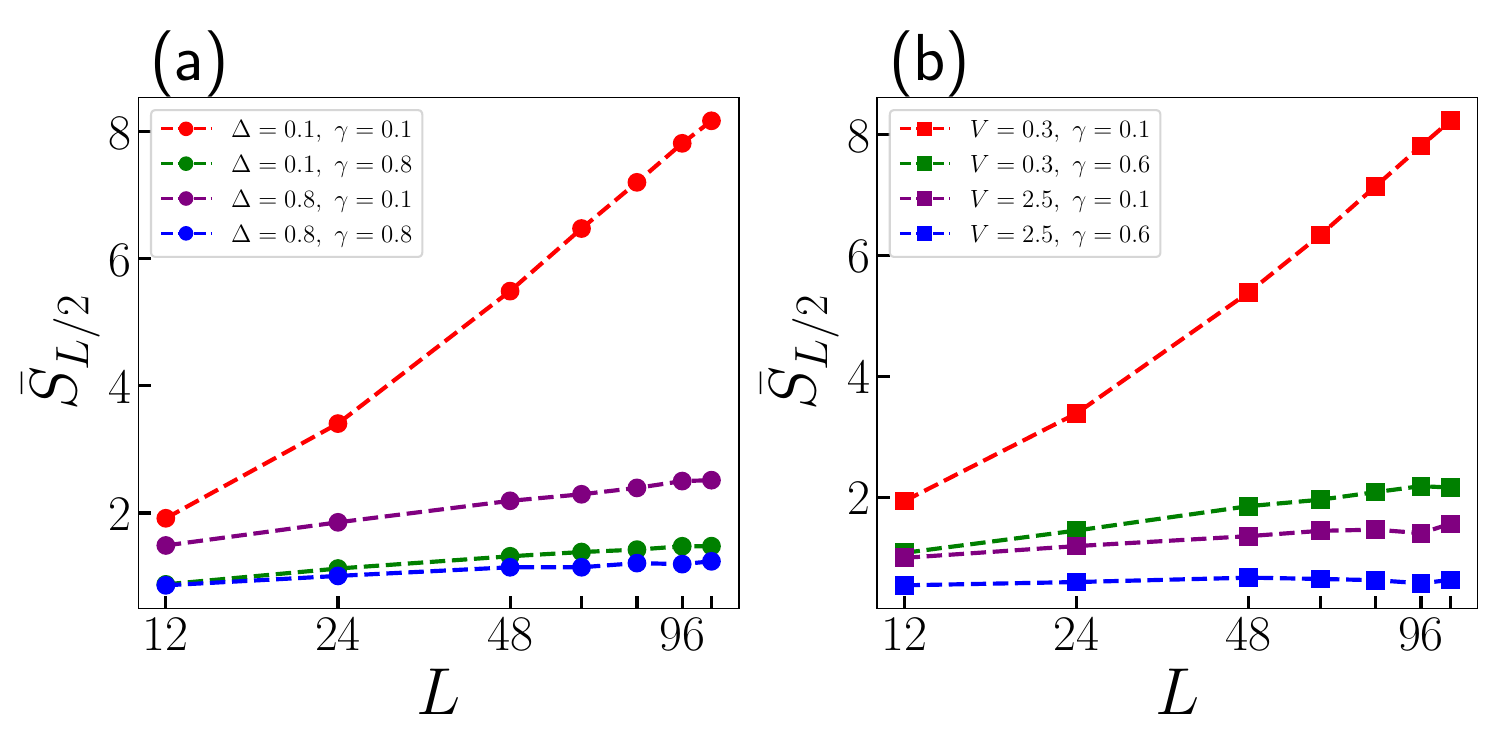} 
\caption{Entanglement entropy versus system size for representative points.  
(a) Results corresponding to the four selected points from Fig.~1(a): \(\Delta=0.1, \gamma=0.1\), \(\Delta=0.1, \gamma=0.8\), \(\Delta=0.8, \gamma=0.1\), and \(\Delta=0.8, \gamma=0.8\);  
(b) Results corresponding to the four selected points from Fig.~1(b): \(V=0.3, \gamma=0.1\), \(V=0.3, \gamma=0.6\), \(V=2.5, \gamma=0.1\), and \(V=2.5, \gamma=0.6\).  
The plots show the half-chain entanglement entropy \(\bar{S}_{L/2}\) as a function of system size \(L\), revealing the scaling behavior under different potential and measurement strengths.  Log-log axes are used throughout for visual clarity. 
}\label{phasedot}
\end{figure}
\begin{figure}[bt]
\hspace*{-0.5\textwidth}
\includegraphics[width=0.5\textwidth]{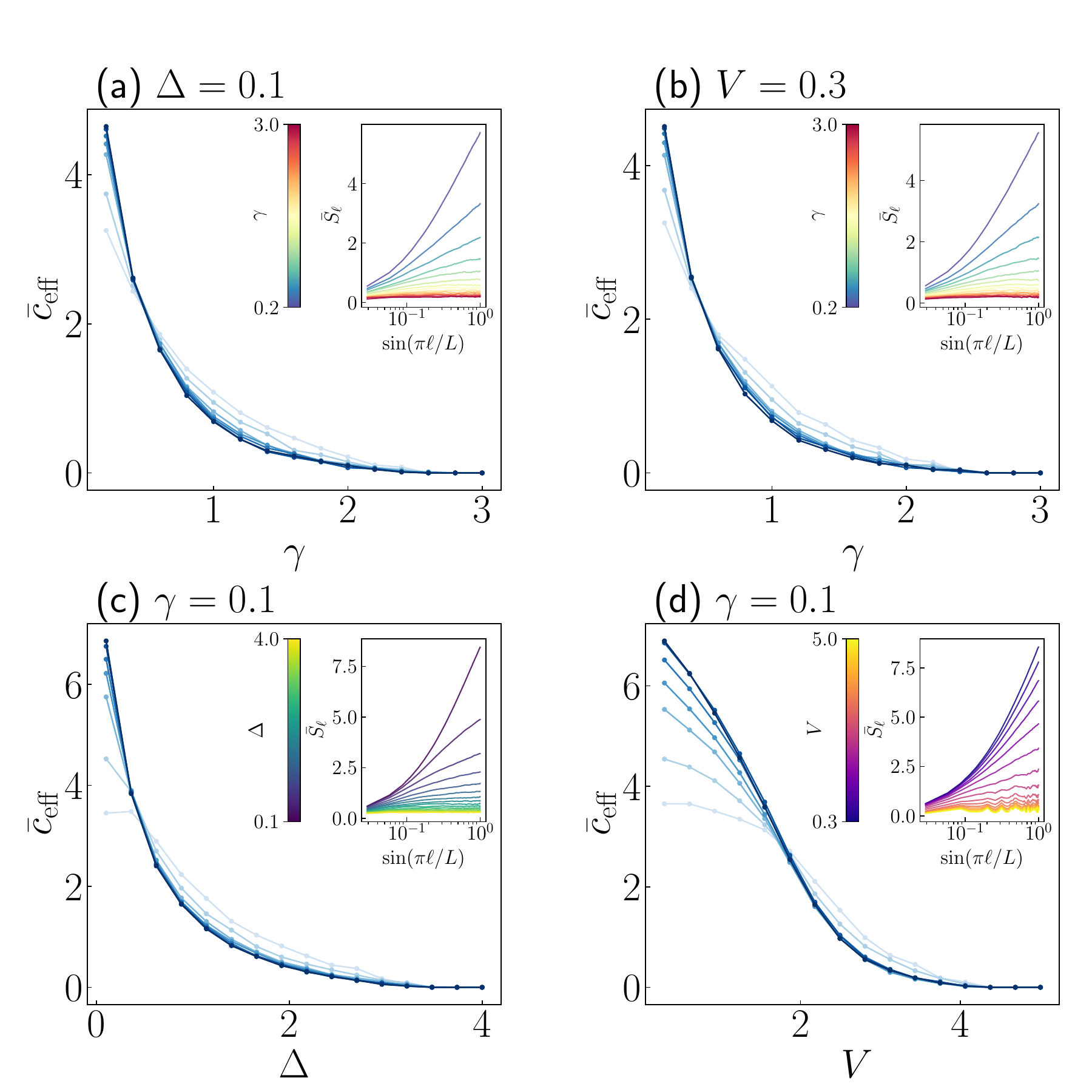} 
\caption{Effective central charge \(\bar{c}_{\mathrm{eff}}\) versus various control parameters with entanglement scaling insets. (a) Fixed \(\Delta = 0.1\), varying measurement strength ; (b) fixed \(V = 0.3\), varying measurement strength ; (c) fixed \(\gamma = 0.1\), varying SP strength ; (d) fixed \(\gamma = 0.1\), varying QPP strength. Insets display entanglement entropy versus \(\sin(\pi \ell / L)\) on a logarithmic \(x\)-axis to emphasize scaling behavior.  Shades of blue from light to dark correspond to system sizes \(L = 12, 24, 48, 64, 80, 96, 112\), the variation in system size is also reflected in the extent of the x-axis in the insets.}\label{S4}
\end{figure}
As stated in Sec.~\ref{sec2}, in the specific study, we divide the system into two subsystems, $A$ and $B$, in order to calculate the entanglement entropy. Consider such a bipartite system governed by purely quantum dynamics, the growth rate of its entanglement entropy takes the following form~\cite{VanAcoleyen_2013_PRL,Bennett_2003_IEEETransIT,Bravyi_2007_PhysRevA,minato,Skinner2019}:
\begin{align}
\dot{S}_{\ell} & =-i\left\|H_{A B}\right\| \lambda(\rho) ,\\
\lambda(\rho) & :=\operatorname{Tr}\left(h_{A B}\left[\rho, \rho_A \otimes \mathbf{I}_B\right]\right),\label{sg}
\end{align}
where $\rho=|\psi(t)\rangle\langle\psi(t)|, h_{A B}:=H_{A B} /\left\|H_{A B}\right\|(\|\ldots\|$ is the operator norm), and $\mathbf{I}_B$ is the identity operator for subsystem $B$. The Hamiltonian $H_{A B}$ denotes the boundary interaction between subsystems $A$ and $B$~\cite{VanAcoleyen_2013_PRL} :
\begin{equation}
H_{A B}=\sum_{i \in A} \sum_{j \in B} h_{i, j},
\end{equation}
where $h_{i, j}$ is an interaction operator acting on sites $i$ and $j$. Of interest is the case where $\ell=L / 2$. In our study of a fermionic system without long-range interactions, the entanglement entropy growth rate is finite and takes a constant value, denoted as \( G_0 \). Since measurements and localized  potentials both act to disentangle the system, even when small, the original volume-law entanglement entropy growth 
\( S \sim G_{0} t \) is transformed, under their combined effect, 
into a logarithmic form \( S \sim \log(t) \), and, upon further increase, 
crosses over to an area-law scaling \( S \sim t^{0} \). 
In our study, the phase boundary corresponds to the critical line separating 
the log-law region from the area-law region. 
Therefore, we find that at the phase boundary the entanglement entropy growth follows 
\( S \sim C_{0} \log(t) \), representing the lower bound of the log-law region 
and the upper bound of the area-law region (see Appendix.~\ref{A1}), 
where \( C_{0} \) is a small constant. Assuming that under the influence of measurements and localized  potentials the initial entanglement growth rate is renormalized to $\Gamma(\gamma,P)G_0,$ then at criticality one can write:
\begin{equation}
\Gamma(\gamma)+\Gamma(P)\sim C_0,\label{ss1}
\end{equation}
$\Gamma(\gamma)$ represents the entanglement suppression rate induced by the measurement strength \(\gamma\). It quantifies the degree to which entanglement growth is suppressed per unit time under continuous measurement interventions. When \(\gamma\) is sufficiently large, entanglement growth can be strongly inhibited or even fully halted. To analyze this process quantitatively, we decompose the system's evolution into individual quantum trajectories. Each trajectory corresponds to a specific measurement history, and the dynamics are governed by an effective non-Hermitian Hamiltonian~\cite{Alberton2021}:
\begin{equation}
H_{\mathrm{eff}}=H-i \frac{\gamma}{2} \sum_u\mathcal{L}_u^{\dagger}\mathcal{L}_u,
\label{eq:nonherm}
\end{equation}
where \(H\) is the original Hermitian Hamiltonian, \(\gamma\) is the measurement rate, and \(\mathcal{L}_u\) is the Lindblad operator. The non-Hermitian~\cite{Yang2025ReversingSkin,
Yang2025BeyondSymmetry,
Liu2024SIEC,
Qin2025ManyBodySkin,
Koh2025EdgeClusterBursts,
Shen2025HyperbolicSkin,
Yao2018EdgeStates,
Yao2018ChernBands,
Song2019ChiralDamping,
Song2019RealSpaceInvariant,
shen2025observation,
qin2025dynamical,
Xiao2020BulkBoundary,
yang2025non} term captures the non-unitary dissipative effects introduced by measurements.  Under this non-Hermitian evolution, Under the trajectory-weighted average, the initial entanglement-entropy growth rate of the system in the presence of measurements is renormalized to (please refer to Appendix.\ref{A1} for details):

\begin{equation}
\Gamma(\gamma)\sim\;G_0\exp(-\gamma).
\label{ss2}
\end{equation}

As shown in Eq.~\eqref{sg}, under purely unitary evolution, entanglement growth is driven by the boundary coupling between subsystems \(A\) and \(B\). The entanglement growth rate is fundamentally determined by the Hamiltonian term \(H_{AB}\) that connects \(A\) and \(B\). In the presence of a localized  potential, the spatial localization of particle states significantly suppresses the boundary coupling between regions, leading to a pronounced reduction in the entanglement growth rate~\cite{Bardarson2012Entanglement}. Corresponding to the distinct localization properties of the wave functions in the SP and the QPP systems, the entanglement growth rate is modified by the localized  potentials as:

\begin{equation}
\Gamma(P) \;\sim\; \begin{cases}G_0\exp (-\,\,\Delta^2) ,& \text { SP  } \\ G_0\exp (-\,\,V), & \text { QPP }\end{cases}\label{ss3}
\end{equation}
where \(G_{0}\) is the reference growth rate in the absence of any potential,  $\Delta$ is the SP strength.  Based on Eq.~\eqref{ss1} and the envelope function of the entanglement growth rate Eq.~\eqref{ss2} and Eq.~\eqref{ss3}, the phase boundaries in the systems with SP or QPP can be written as:
\begin{equation}
C_0=\begin{cases}\exp(-6\gamma^{3/2})+\exp (-\,8\,\Delta^2),& \text { SP  }\\\exp(-6\gamma^{3/2})+\exp (-\,V/2). & \text { QPP }\end{cases}\label{fs}
\end{equation}
Eq.\eqref{fs} provides a phenomenological envelope-function expression~\cite{Gluck_2002_PhysRep}, and in the Anderson case~\cite{Kramer1993}, the measurement further introduces a super-exponential envelope that strongly suppresses the coefficient $C_{0}$. To validate Eq.~\eqref{fs}, we first map the effective central charge $\bar{c}_{\mathrm{eff}}$ over the parameter space spanned by the measurement strength $\gamma$ and the localized  potential strength (either Stark, $\Delta$, or quasi-periodic, $V$), thereby characterizing the scaling regimes of the entanglement entropy. As shown in Fig.~\ref{phase}, the phase diagrams under varying measurement and potential strengths are presented for the SP and QPP systems. The colormap encodes $\bar{c}_{\mathrm{eff}}$ extracted from half-chain entanglement entropy; the horizontal axis is the measurement rate $\gamma$, and the vertical axis is the potential strength $\Delta$ or $V$. The diagrams clearly separate area-law and log-law regimes: regions with large $\bar{c}_{\mathrm{eff}}$ correspond to log-law growth (an entanglement-spreading phase), whereas regions with $\bar{c}_{\mathrm{eff}} \approx 0$ exhibit suppressed entanglement.
\begin{figure}[bt]
\hspace*{-0.5\textwidth}
\includegraphics[width=0.5\textwidth]{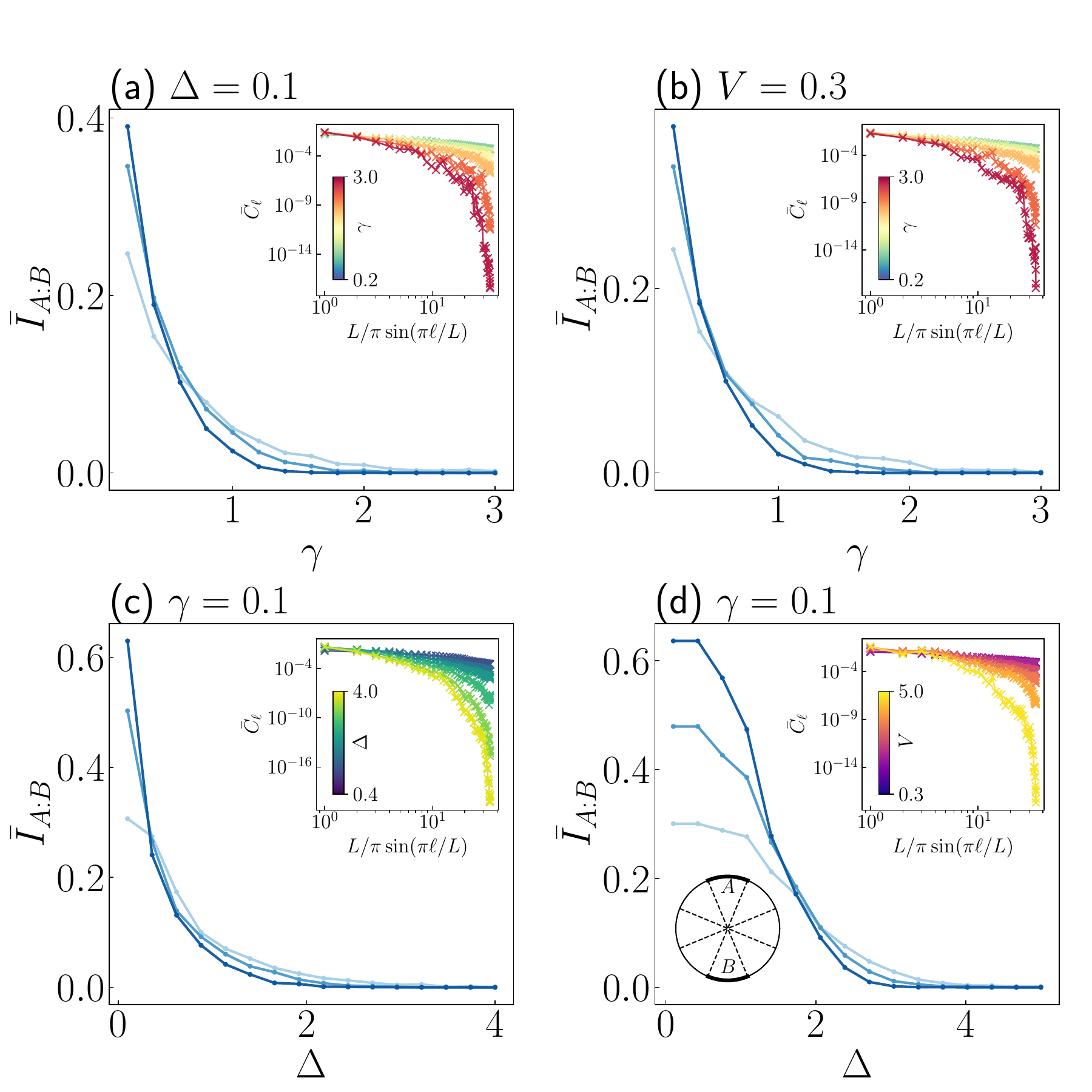} 
\caption{Mutual information under varied control parameters, with scaling insets. (a) Fixed \(\Delta = 0.1\), varying measurement strength ; (b) fixed potential strength \(V = 0.3\), varying measurement strength; (c) fixed measurement strength \(\gamma = 0.1\), varying SP strength; (d) fixed \(\gamma = 0.1\), varying QPP strength. Main panels show mutual information as a function of each parameter. Insets present correlation functions versus \(\sin(\pi \ell / L)\), log-scaled on the \(x\)-axis to highlight scaling transitions. Shades of blue from light to dark correspond to system sizes \(L = 12, 24, 48, 64, 80, 96, 112\), the variation in system size is also reflected in the extent of the x-axis in the insets.
}\label{ICL}
\end{figure}
We find that increasing either the SP or QPP shrinks the entangled region and eventually eliminates the phase boundary, although the deformation is qualitatively different between SP and QPP. The dashed curves are fits to Eq.~\eqref{fs}, which agree closely with critical points obtained via finite-size scaling (black dots). As the potential strength increases, the entangled region narrows, the boundary bends, and eventually terminates, reflecting the progressively stronger suppression of entanglement propagation caused by localization. Importantly, the boundary deforms differently in the two cases: for SP, it changes super-exponentially with $\Delta^2$; for QPP, it changes exponentially with $V$ results that are fully consistent with the analytical structure established earlier. Together, they reveal a measurement-localization coupled entanglement phase transition: when the localization potential is small, the phase boundary is primarily determined by the measurement strength, and its descent is relatively gradual; as the localization potential increases, its suppression of entanglement growth causes the boundary to drop rapidly. However, because the Stark localization induced by SP is stronger than the Anderson localization induced by QPP, the phase boundary in Fig.~\ref{phase}(a) falls faster, reaching the area-law region much earlier. It is worth noting that, in the limit of vanishing localized potential $V(\Delta) = 0$, the system can be mapped onto a nonlinear $\sigma$ model, in which no entanglement phase transition occurs~\cite{Igo_prx}. In the limit $\gamma = 0$, in the thermodynamic limit, any finite strength of the SP or disorder ($W$) 
will lead to localization~\cite{Gluck_2002_PhysRep,Mendez1993,Anderson1958,Kramer1993}, suppressing long-range information propagation, resulting in an area-law state, and no phase transition occurs in the system. However, due to finite-size effects, the localization length in this case exceeds the system size, 
which causes the $\gamma = 0$ region in Fig.~\ref{phase}(a) and Fig.~\ref{Dphase} to still exhibit signatures of a phase transition. 
For the QPP, Anderson localization occurs only when $V > 2J$~\cite{Aubry1980AIPS}, which is consistent with the results at $\gamma = 0$ in Fig.~\ref{phase}(b).

To further verify the scaling characteristics of different regions, we examine the dependence of the entanglement entropy $\bar{S}$ on the system size $L$, selecting representative points (red, green, blue, purple) highlighted in Fig.~\ref{phase}. The corresponding results are shown in Fig.~\ref{phasedot}. It is evident that, under different measurement strengths and localization strengths, the dependence of $\bar{S}$ on $L$ exhibits significant differences. At the red-marked point in the phase diagram within the log-law region,$\bar{S}$ shows a clear linear growth with $\log(L)$, indicating that the system is in a log-law-dominated phase. In this regime, the measurement rate and localized  potential are insufficient to significantly disrupt the propagation of particles or information across the entire system, leading to logarithmic growth of entanglement. In contrast, in the area-law region (i.e., at non-red points), $\bar{S}$ rapidly saturates to a constant as $\log(L)$ increases. This reflects the fact that, under strong measurement or strong localization, the propagation of quantum information is strongly suppressed, and entanglement can only be established within a finite range near the boundaries between adjacent regions, without extending throughout the system. This transition from a log-law to an area-law scaling reflects a measurement-and-localization driven entanglement phase transition: when the measurement or localization strength crosses a critical value, the system's information propagation capability undergoes a sudden change, leading to an abrupt shift in the scaling behavior of the entanglement entropy and the emergence of a well-defined critical boundary.

\subsection{Phases and phase transitions }
We next follow the finite-size flow of the effective central
charge \(\bar{c}_{\mathrm{eff}}\) under OBCs. For each size \(L\) we fit the entanglement profile to Eq.~\eqref{eq:cft_scaling} and obtain a single \(\bar{c}_{\mathrm{eff}}\)  subjected to different measurement intensities and localized  potentials, as illustrated in Fig.~\ref{S4}. Near criticality, the half-chain entanglement entropy exhibits logarithmic scaling consistent with non-unitary conformal field theory~\cite{Calabrese2004,Calabrese2009}:
\begin{equation}
\bar{S}(\ell, L) = \frac{\bar{c}_{\mathrm{eff}}}{3} \ln\left[ \frac{L}{\pi} \sin\left( \frac{\pi \ell}{L} \right) \right] + s_0,
\label{eq:cft_scaling}
\end{equation}
where \(\ell\) is the subsystem size, \(\bar{c}_{\mathrm{eff}}\) is the effective central charge, and \(s_0\) is a non-universal constant. In the thermodynamic limit, \(\bar{c}_{\mathrm{eff}} \to 0\) in the area-law phase, while it remains finite in the logarithmic phase, thus serving as a diagnostic for the transition~\cite{Alberton2021,Buchhold2021}. As shown in Fig.~\ref{S4}(a,b), when the localized  potential is fixed at a small value ($\Delta = 0.1$ or $V = 0.3$), increasing the measurement strength $\gamma$ causes $\bar{c}_{\mathrm{eff}}$ curves for different system sizes to gradually intersect, indicating a transition near $\gamma \approx 0.3$. The insets strongly corroborate this behavior: as $\gamma$ increases, the entanglement entropy transitions from rapid growth to saturation, signaling a crossover from log-law to area-law scaling. Similarly, in Fig.~\ref{S4}(c,d), when the measurement strength is fixed at a low value ($\gamma = 0.1$), increasing the localized  potential strength leads to similar crossing behavior among different system sizes, with transition points occurring around $\Delta \approx 0.3$ and $V \approx 1.8$, respectively. The insets in Fig.~\ref{S4} clearly demonstrate this effect: the entanglement entropy drops from steep growth to a nearly constant value as localization increases, again signifying a transition from log-law to area-law behavior. These results collectively indicate that both measurement strength and localized  potential can induce a phase transition in the system. In both cases, entanglement growth is suppressed, although the mechanisms and suppression profiles differ between SP and QPP. These observations reveal how both measurement and localization reduce boundary coupling and suppress entanglement growth. The distinct functional behaviors of SP and QPP reinforce the validity of \(\bar{c}_{\mathrm{eff}}\) as a transition indicator.

Another independent indicator of the entanglement phase transition is the mutual information. As illustrated in Fig.~\ref{ICL}(d), the system is partitioned into four segments: subsystems $A$ and $B$, each occupying $L/8$ sites, are separated by two buffer regions of length $3L/8$. The mutual information is defined as
\begin{equation}
\bar{I}_{A:B} = \bar{S}_A + \bar{S}_B - \bar{S}_{AB},
\end{equation}
where $\bar{S}_{A}$, $\bar{S}_{B}$, and $\bar{S}_{AB}$ denote the entanglement entropies of regions $A$, $B$, and $A \cup B$, respectively. Mutual information has proven to be a sensitive diagnostic of MIPT in a variety of systems~\cite{ Li2019,Fuji2020}, due to its ability to capture nonlocal correlations. In our study, we adopt the same set of parameters used for computing $\bar{c}_{\mathrm{eff}}$. As shown in Fig.~\ref{ICL}(a,b), when the localized  potential is set to a small value ($\Delta = 0.1$ or $V = 0.3$), increasing the measurement strength $\gamma$ leads to a crossing in mutual information curves across different system sizes, signaling the onset of a transition. Similarly, in Fig.~\ref{ICL}(c,d), fixing the measurement strength and gradually increasing the potential strength ($\Delta$ or $V$) also induces such crossings, indicative of a phase boundary. To further understand the origin of these behaviors, we compute the connected correlation function defined as
\begin{equation}
\bar{C}_\ell = \langle \hat{n}_{L/2} \rangle \langle \hat{n}_{L/2+\ell} \rangle - \langle \hat{n}_{L/2} \hat{n}_{L/2+\ell} \rangle,
\end{equation}
which reduces to $\bar{C}_\ell = |\langle \hat{c}^\dagger_{L/2} \hat{c}_{L/2+\ell} \rangle|^2$ for the Slater-determinant state $|\psi(t)\rangle$. As shown in the insets of Fig.~\ref{ICL}, both increasing measurement strength and localized  potential lead to qualitatively similar changes in the correlation structure. In the low measurement or weak localization regime, $\bar{C}_\ell$ decays algebraically, $\bar{C}_\ell \sim \ell^{-\eta}$, indicating long-range correlations and rapid entanglement growth. In the strong measurement or deep localized regime, $\bar{C}_\ell$ decays exponentially, $\bar{C}_\ell \sim \exp(-\ell/\xi)$, suggesting short-range correlations and saturation of entanglement entropy consistent with area-law behavior. This sharp decay in the correlation function confirms a transition from an entangled phase with extended quantum coherence to a localized phase dominated by measurement or potential-induced suppression. Together with mutual information crossings, these results demonstrate that both measurement and disorder drive a phase transition that limits entanglement growth, with distinct suppression mechanisms depending on the type of localized  potential applied.
\begin{figure}[bt]
\hspace*{-0.5\textwidth}
\includegraphics[width=0.5\textwidth]{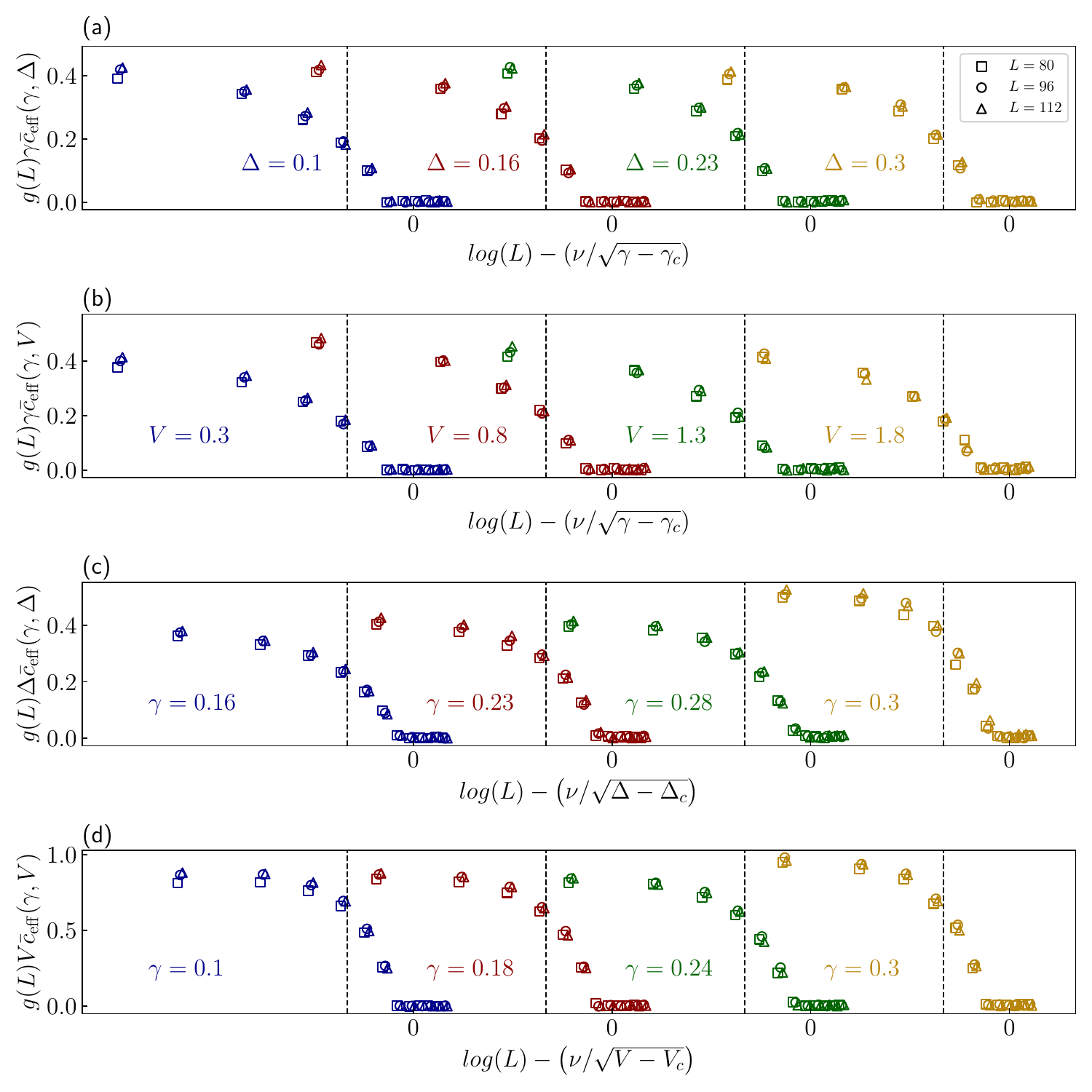} 
\caption{Finite-size scaling for the effective central charge with the BKT scenario, (a): $g(L) \gamma \bar{c}_{\rm eff}(\gamma, \Delta)$ versus $\log(L)-\nu / \sqrt{\gamma-\gamma_c}$, (b) :  $g(L) \gamma \bar{c}_{\rm eff}(\gamma, V)$ versus $\log L-\nu / \sqrt{\gamma-\gamma_c}$ .(c): $g(L) \Delta \bar{c}_{\rm eff}(\gamma, \Delta)$ versus $\log(L)-\nu / \sqrt{\Delta-\Delta_c}$, (d) :  $g(L) V \bar{c}_{\rm eff}(\gamma, V)$ versus $\log L-\nu / \sqrt{V-V_c}$. $g(L) = [1 + 1/(2 \log L - \alpha)]^{-1}$. Detailed numerical data are provided in Table.~\ref{table1}.
}\label{DC}
\end{figure}
To quantitatively determine the location and nature of the entanglement phase transition, we perform a finite-size scaling and data collapse analysis of the effective entanglement quantity \(\bar{c}_{\mathrm{eff}}\) within the Berezinskii-Kosterlitz-Thouless (BKT) framework. Given the pronounced finite-size corrections in free-fermion circuits, simple visual identification of crossing points is unreliable. We adopt a single-parameter scaling ansatz with a mild size correction factor \(g(L)\), incorporating the essential singularity of the BKT correlation length, expressed as~\cite{Alberton2021,harada1997universal} 
\begin{equation}
g(L)\,x\,\bar{c}_{\mathrm{eff}}(x,y) = F[\log(L\,\xi(x,x_c(y)))], \label{dceq}
\end{equation}
\begin{figure}[bt]
\hspace*{-0.5\textwidth}
\includegraphics[width=0.5\textwidth]{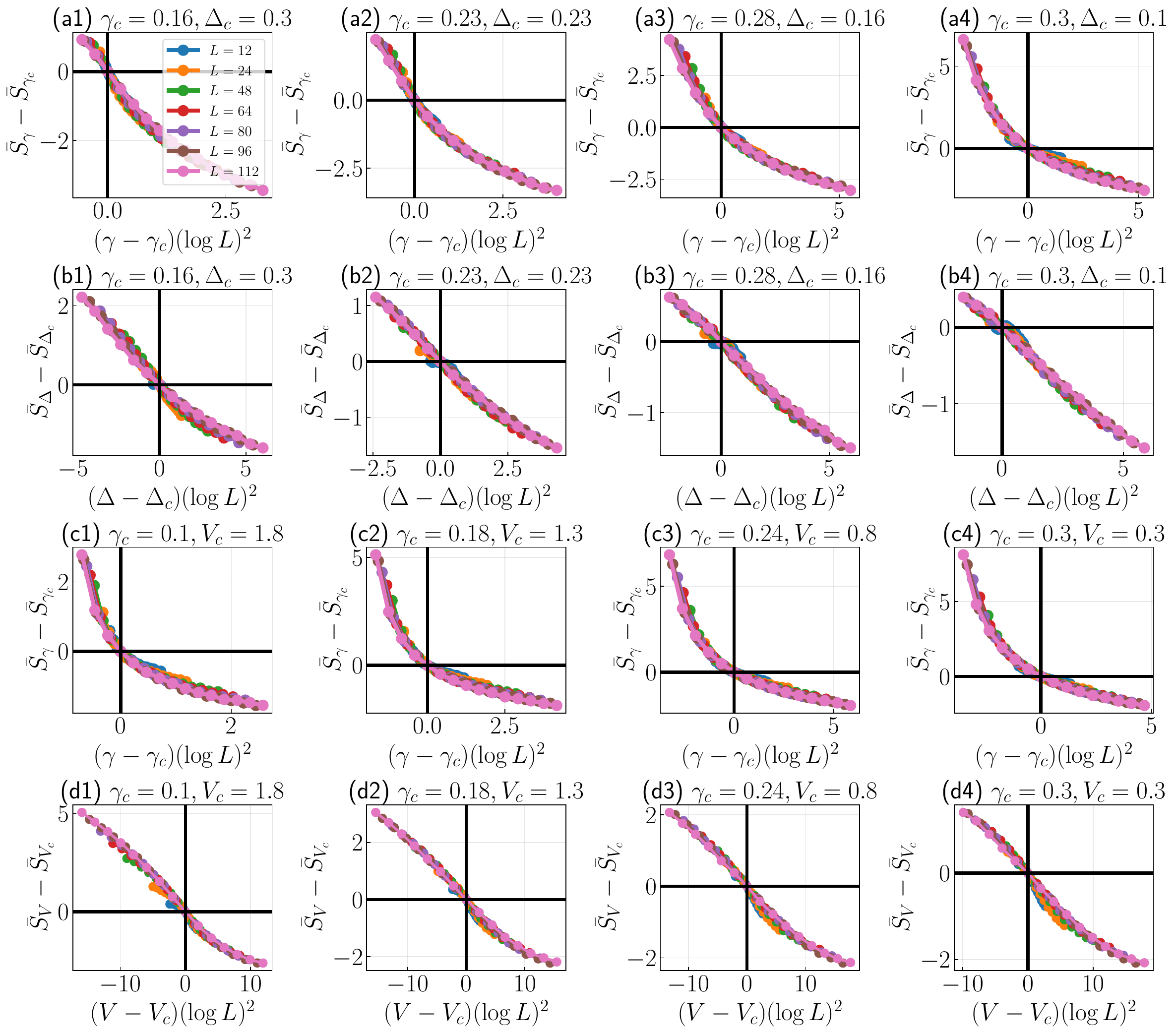} 
\caption{The finite-size scaling collapse of the entanglement entropy is performed under the assumption of BKT scaling 
of the correlation length. Subfigures (a) and (b) correspond respectively to the critical points 
$(\gamma_c, \Delta_c) = (0.16, 0.3), (0.23, 0.23), (0.28, 0.16), (0.30, 0.10)$, 
where panel (a) is obtained from the collapse of $\gamma$ data and panel (b) from the collapse of $\Delta$ data. 
Subfigures (c) and (d) correspond respectively to the critical points 
$(\gamma_c, V_c) = (0.10, 1.8), (0.18, 1.3), (0.24, 1.8), (0.30, 0.3)$, 
where panel (c) is obtained from the collapse of $\gamma$ data and panel (d) from the collapse of $V$ data. 
The analyses are performed using system sizes $L = 12, 24, 48, \ldots, 112$.
}\label{BKT}
\end{figure}
\begin{table}[bt]
\caption{\label{table1} Critical exponents and critical points.}
\begin{ruledtabular}
\begin{tabular}{cccc}
$\Delta$ & $\gamma_c$ & $\nu$ & $\alpha$ \\
\hline
0.10 & $0.30 \pm 0.035$ & $3.8 \pm 0.2$ & $5.9 \pm 0.3$ \\
0.16 & $0.28 \pm 0.035$ & $4.1 \pm 0.2$ & $6.2 \pm 0.3$ \\
0.23 & $0.23 \pm 0.045$ & $5.7 \pm 0.3$ & $6.5 \pm 0.45$ \\
0.30 & $0.16 \pm 0.040$ & $6.5 \pm 0.3$ & $7.3 \pm 0.5$ \\
\end{tabular}
\begin{tabular}{cccc}
$V$ & $\gamma_c$ & $\nu$ & $\alpha$ \\
\hline
0.30 & $0.30 \pm 0.04$ & $2.7 \pm 0.3$ & $4.1 \pm 0.35$ \\
0.80 & $0.24 \pm 0.04$ & $2.9 \pm 0.35$ & $4.3 \pm 0.35$ \\
1.30 & $0.18 \pm 0.04$ & $3.3 \pm 0.35$ & $4.5 \pm 0.4$ \\
1.80 & $0.10 \pm 0.045$ & $3.8 \pm 0.4$ & $4.9 \pm 0.4$ \\
\end{tabular}
\begin{tabular}{cccc}
$\gamma(\Delta)$ & $\Delta_c$ & $\nu$ & $\alpha$ \\
\hline
0.16 & $0.30 \pm 0.015$ & $6.5 \pm 0.1$ & $8.9 \pm  0.1$ \\
0.23 & $0.23 \pm 0.03$  & $4.9 \pm 0.2$ & $7.4 \pm 0.3$ \\
0.28 & $0.16 \pm 0.035$ & $4.1 \pm 0.3$ & $6.5 \pm 0.4$ \\
0.30 & $0.10 \pm 0.035$ & $3.3 \pm 0.3$ & $4.9 \pm 0.4$ \\
\end{tabular}
\begin{tabular}{cccc}
$\gamma(V)$ & $V_c$ & $\nu$ & $\alpha$ \\
\hline
0.10 & $1.8 \pm 0.35$ & $3.6 \pm 0.35$ & $5.3 \pm 0.45$ \\
0.18 & $1.3 \pm 0.3$  & $3.2 \pm 0.3$ & $4.8 \pm 0.45$ \\
0.24 & $0.8 \pm 0.15$ & $2.9 \pm 0.2$ & $4.1 \pm 0.3$ \\
0.30 & $0.1 \pm 0.28$ & $2.6 \pm 0.3$ & $3.2 \pm 0.4$ \\
\end{tabular}
\end{ruledtabular}
\end{table}
where \(x \in \{\gamma,\Delta,V\}\) and correlation length \(\xi(x,x_c) \sim \exp[-\nu/\sqrt{|x - x_c|}]\). This yields natural scaling variables \(X_x = \log L - \nu/\sqrt{|x - x_c|}\) and \(Y_x = g(L)\,x\,\bar{c}_{\mathrm{eff}}\), allowing us to absorb leading finite-size drift and achieve data collapse onto a unified master curve across system sizes and parameter paths. Fig.~\ref{DC} illustrates this systematic validation across multiple paths: (a) for fixed localized  potential \(\Delta = (0.1, 0.16, 0.23, 0.3)\), data from system sizes \(L = 80, 96, 112\) collapse well after rescaling, enabling precise determination of \(\gamma_c(\Delta)\) and a shared \(\nu\); (b) repeating the analysis for different potential strengths \(V = \{0.3, 0.8, 1.3, 1.8\}\), we trace the evolution of \(\gamma_c(V)\); (c) scanning \(\Delta\) at fixed measurement strengths \(\gamma = \{0.16, 0.23, 0.28, 0.30\}\), we extract \(\Delta_c(\gamma)\), which agrees with the inverse function from (a); (d) scanning \(V\) at fixed \(\gamma = \{0.10, 0.18, 0.24, 0.30\}\), we determine \(V_c(\gamma)\), completing the critical boundary in the \((\gamma,V)\) plane. These appropriate values of $\gamma_c$ or $\Delta_c ,V_c$ lead to good scaling collapses, which are consistent with the scaling formula Eq.~\ref{dceq}. To more precisely determine the phase transition points and to examine whether the transition belongs to the BKT universality class, we perform finite-size scaling analyses for the four critical points using the following scaling relations~\cite{harada1997universal}:
\begin{equation}
\begin{aligned}
\bar{S}_{\gamma} - \bar{S}_{\gamma_c} &= F\!\left[(\gamma - \gamma_c)(\log L)^2\right], \\
\bar{S}_{\Delta} - \bar{S}_{\Delta_c} &= F\!\left[(\Delta - \Delta_c)(\log L)^2\right],\\
\bar{S}_{V} - \bar{S}_{V_c} &= F\!\left[(V - V_c)(\log L)^2\right],
\end{aligned}
\label{bkteq}
\end{equation}
where $F$ denotes the scaling function. This scaling formula is based on the assumption that the MIPT belongs to a universality class similar to the BKT one~\cite{Alberton2021}, where the correlation length $\xi$ diverges exponentially around the transition point; specifically, 
\begin{equation}
\begin{aligned}
\log \xi &\sim \frac{1}{\sqrt{\gamma - \gamma_c}} \quad &&\text{as } \gamma \to \gamma_c+0, \\
\log \xi &\sim \frac{1}{\sqrt{\Delta - \Delta_c}} \quad &&\text{as } \Delta \to \Delta_c+0, \\
\log \xi &\sim \frac{1}{\sqrt{V - V_c}} \quad &&\text{as } V \to V_c+0.
\end{aligned}
\end{equation}
As shown in Fig.~\ref{BKT}, these transition points yield good scaling collapses, consistent with Eq.~\ref{bkteq}. The successful scaling collapses provide strong evidence that the phase transition, jointly driven by the measurement strength $\gamma$ and the Stark potential strength $\Delta$, falls into the BKT universality class. To further determine the error of the phase transition point , we minimize the cost function~\cite{Szyniszewski2023PRBB54309}, where the error bar of $\gamma_c$ or $\Delta_c ,V_c$ are estimated from the range within which the cost function does not exceed twice their minimum value. The obtained results are presented as circles with error bars in Fig.~\ref{phase} 
and are summarized in Table.~\ref{table1}. The critical points obtained from all four paths are consistently described using a unified scaling function \(F\) and a common critical exponent \(\nu\), indicating that the localized  potential does not merely shift an isolated critical point but cooperates with measurement strength to generate a continuous phase boundary in the \((\gamma,\Delta,V)\) parameter space. Mathematically, while measurement tends to suppress entanglement, the localized  potential restructures the state and alters information transport, and the competition between these effects determines the geometry and location of the phase boundary governed by the BKT singularity. The successful data collapses validate the applicability and robustness of the BKT scaling framework and provide a quantitative foundation for extracting the critical points \(\gamma_c(\Delta)\), \(\gamma_c(V)\), \(\Delta_c(\gamma)\), \(V_c(\gamma)\), and the correlation length exponent \(\nu\).

\section{conclusion and discussion}\label{sec4}
We investigated the continuous measurement of free fermions under  SP and QPP. Our study shows that when the SP strength or the measurement strength is weak, the system remains in a stable logarithmic-scaling phase. Moreover, we found that increasing either the measurement strength or the SP triggers a phase transition, pushing the system into an area-law phase. When the potential strength is strong, we demonstrate that the MIPT disappears, and regardless of the measurement strength, the system remains stable in the area-law phase. To further support these results, we analyzed the central charge, mutual information, and correlation functions. The results reveal that using central charge and mutual information to characterize the system's entanglement behavior clearly exposes a crossover phenomenon, indicating that an increase in either measurement strength or SP leads to an entanglement phase transition. These findings suggest that, qualitatively, Stark localization (caused by SP) and Anderson localization (caused by QPP) play equivalent roles in driving entanglement phase transitions, and both belong to the BKT universality class. 

Furthermore, we construct a phase diagram under different localization mechanisms, which diverges at a boundary characterized by entanglement propagation. We find that the monitored Stark-localized system decays rapidly under weak measurement strength, whereas the Anderson-localized system exhibits a slower decay.

Our work highlights the distinctive behavior of MIPTs and relaxation dynamics in Stark-driven monitored systems, and provides a perspective on the characterization of the entanglement phase boundary arising from the combined effects of measurement and localization.
\begin{acknowledgments}
We thank Shuo Liu and Ze-Chuan Liu for the discussions. J.-X. Zhong acknowledges the National Natural Science Foundation of China (Grant No.12374046) and the Shanghai Science and Technology Action Plan (Grant No.24LZ1400800). H.-Z. Li is supported by the CSC scholarship.
\end{acknowledgments}
%%%%%%%%%%%%%%%%%%%%%%%%%%%%%%%%%%%%%%%%%%%%%%%%%%%%%%%%%%
%%%%%%%%%%%%%%%%%%%%% APPENDIX %%%%%%%%%%%%%%%%%%%%%%%%%%%
%%%%%%%%%%%%%%%%%%%%%%%%%%%%%%%%%%%%%%%%%%%%%%%%%%%%%%%%%%
\appendix
\section{Discussion on the Phase Boundary}\label{A1}
\begin{figure}[bt]
\hspace*{-0.5\textwidth}
\includegraphics[width=0.5\textwidth]{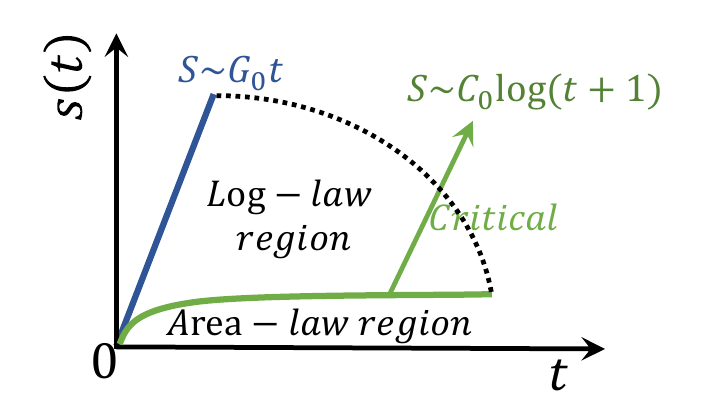} 
\caption{The growth of bipartite entanglement entropy between the two semi-infinite halves of an infinite chain is shown. In the entangling phase
(upper curve) the entanglement grows ballistically with time. At the critical point ( middle curve), the entanglement grows logarithmically. In the disentangling phase ( lower curve), the entanglement saturates to a finite value.}\label{fig10}
\end{figure}

As shown in Fig.~\ref{fig10}, the entanglement entropy growth of the system exhibits distinct behaviors in different phases. From the figure, in the initial moment ($t \to 0$) within the Entangling phase the growth obeys:
\begin{equation}
S(t) = G_0\,t,
\end{equation}
whereas under Critical dynamics it follows:
\begin{equation}
S(t) = C_0\log\bigl(t+1\bigr).
\end{equation}
After introducing measurements or localized  potentials, the entanglement growth of the system is modified.  Denoting the corrections due to measurements and localized  potentials by $\Gamma(\gamma)$ and $\Gamma(P)$, respectively, we have:  
\begin{equation}
S(t)=\Gamma(\gamma)G_0t,S(t)=\Gamma(P)G_0t.  
\end{equation}
Hence, once measurements and localized  potentials suppress the Entangling-phase growth rate down to that of the critical phase, the system reaches criticality:
\begin{equation}
G_0\Gamma(\gamma)+G_0\Gamma(P)= C_0/(t+1).
\end{equation}
For the measurement-induced renormalization of the entanglement-growth rate $\Gamma(\gamma)$, We know that for a continuously monitored fermionic system, the dynamics is governed by the Lindblad master equation:  
\begin{equation}
\begin{aligned}
\frac{d \rho_t}{d t}&= \mathcal{L} \rho_t \\
&=-i \hat{H}_{\mathrm{eff}} \rho_t+i \rho_t \hat{H}_{\mathrm{eff}}^{\dagger}+\gamma \sum_m^{L-1} \hat{L}_m \rho_t \hat{L}_m^{\dagger},
\end{aligned}\label{eq5}
\end{equation}
where the non-Hermitian effective Hamiltonian \(H_{\mathrm{eff}}\) is given by Eq.~(5).  
In our actual calculations, we employ the It\^{o} stochastic differential equation averaged over multiple trajectories,
so that any amplitude to propagate across \(L\) sites without a jump decays as:
\begin{equation}
p_{\rm surv}(L,t)
=e^{-\gamma\,L\,t}.
\end{equation}
Meanwhile, the unitary Lieb-Robinson bound gives for the creation amplitude across distance \(L\) in time \(t\):
\begin{equation}
A_0(L,t)\;\le\;C_0\,
\exp (-\tfrac{L-v\,t}{\xi_0}).
\end{equation}
Requiring both "no-jump" and unitary spread,
\begin{equation}
A(\gamma;L,t)
\;\lesssim\;
C_0\,\exp (-\tfrac{L-vt}{\xi_0})\,
\exp (-\gamma\,L\,t).
\end{equation}
A saddle-point optimization in \((L,t)\) shows that the maximal amplitude obeys:
\begin{equation}
\max_{L,t}A(\gamma;L,t)\;\propto\;\exp \bigl(-\,a\,\gamma^b\bigr).
\end{equation}
Since \(\Gamma\) is proportional to this amplitude scale, we set
\begin{equation}
\Gamma(\gamma)
=\exp(-\,a\,\gamma^b),
\end{equation}
this is the entanglement suppression rate induced by measurement.
\begin{figure}[bt]
\hspace*{-0.5\textwidth}
\includegraphics[width=0.5\textwidth]{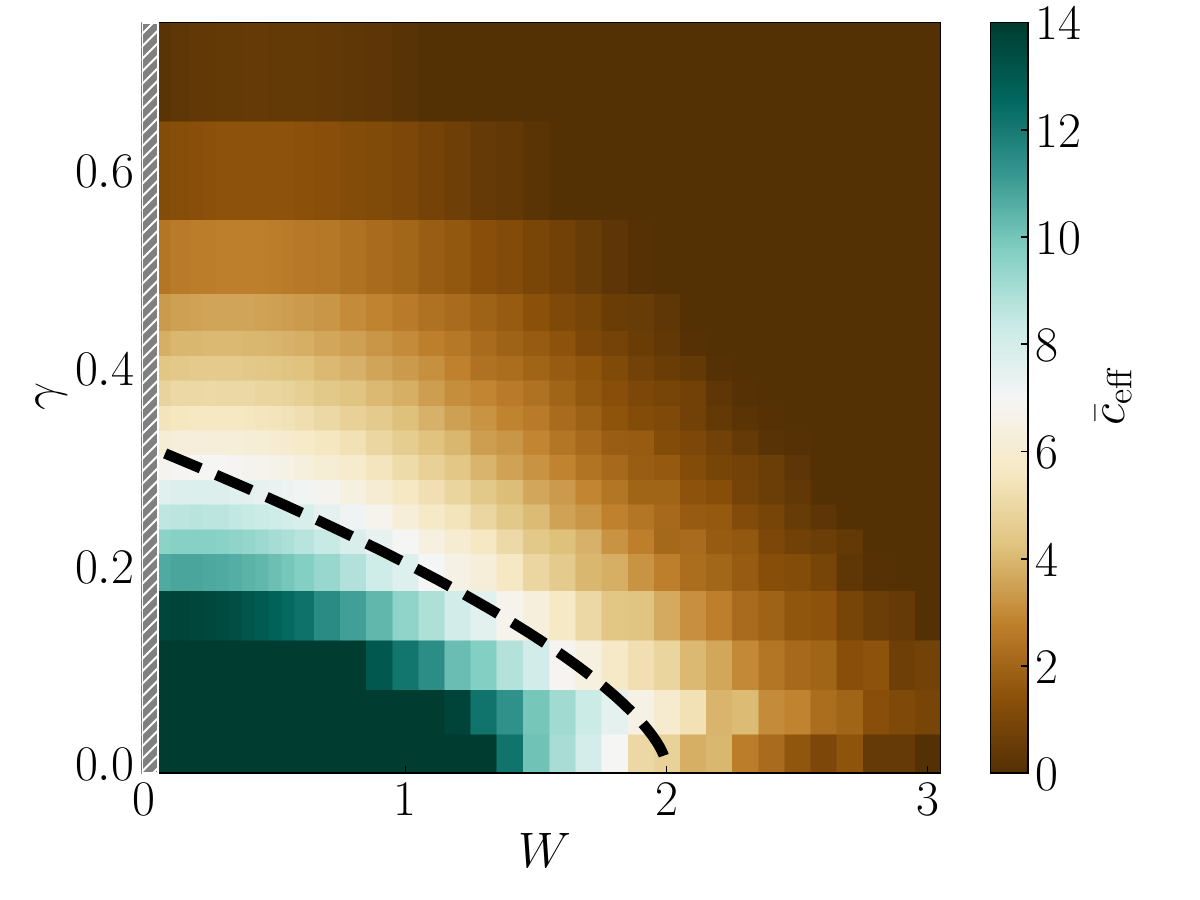} 
\caption{Phase diagrams and phase boundaries extracted from the effective central charge under joint tuning of the measurement strength and Anderson disorder.  
Colors indicate the effective central charge \(\bar{c}_{\mathrm{eff}}\), extracted from the finite-size scaling of half-chain entanglement entropy: green corresponds to large \(\bar{c}_{\mathrm{eff}}\) values (log-law-like regime), while brown corresponds to \(\bar{c}_{\mathrm{eff}} \rightarrow 0\) (area-law regime). 
The grey region corresponds to the case of a very weak localization potential, where we tend to revert to the results obtained from the Keldysh field theory~\cite{Igo_prx}. 
Dashed curves denote the semi-analytical expressions for phase boundaries derived earlier. }\label{Dphase}
\end{figure}
For the localized  potentials induced renormalization of the
entanglement-growth rate $\Gamma(P)$, consider the single-particle tight-binding Hamiltonian as shown in Eq.~\eqref{hami}, as an illustrative example, we choose the SP:
\begin{align}
    P_j =  F_j ,
\end{align}
$F_j = \Delta \cdot j/L$, where $\Delta$ denotes the maximum SP. Its eigenstates-the Wannier-Stark ladder-expand in the site basis as:
\begin{equation}
\psi_n(j)
=J_{\,j-n} \Bigl(\tfrac{2J}{\Delta}\Bigr),
\end{equation}
where \(J_m(z)\) is the Bessel function.  For large order \(m\gg z\), one uses the asymptotic
\begin{align}
J_m(z)&\sim 
\exp (-\,m\ln\tfrac{2m}{e\,z}+O(\ln m))\\
&\approx\;\exp (-\,c\,m^d).
\end{align}
Setting \(m\approx|j-n|\) and defining the localization length \(\xi=2J/\Delta\), the effective boundary coupling between regions \(A\) and \(B\) obeys
\begin{align}
\|H_{AB}(\Delta)\|
&=\sum_{i\in A,j\in B}\bigl|J_{\,i-j}(2J/\Delta)\bigr|\\
\;&\approx\;\|H_{AB}\|\,
\exp (-\,c\,\Delta^d).
\end{align}
By the same Lieb-Robinson argument, the unitary entanglement growth rate is bounded by:
\begin{equation}
\Gamma(\Delta)\;\le\;\|H_{AB}(\Delta)\|,
\end{equation}
and we choose the tight form:
\begin{equation}
\Gamma(\Delta)
=G_0\;\exp \bigl[-\,c\,\Delta^d\bigr],
\end{equation}
this is the entanglement entropy growth rate corrected by the SP. Similarly, for a one-dimensional QPP Eq.~\eqref{hami},  the single-particle eigenfunctions localize exponentially with localization length \(\xi(V)\).  Hence the entanglement entropy growth rate is suppressed to  
\begin{equation}
\Gamma(V)
=G_0\;\exp (-\,e\,V).
\end{equation}
Then, the phase boundaries in the systems with SP or QPP can be written as:
\begin{equation}
C_0=\begin{cases}\exp(-a\gamma^b)+\exp (-\,c\,\Delta^d),& \text { SP  }\\\exp(-a\gamma^b)+\exp (-\,e\,V). & \text { QPP }\end{cases}\label{23}
\end{equation}
Eq.~\eqref{23} is the general solution we finally obtain, and in this paper we fix the parameters as $(a,b,c,d,e,C_0) = (6,1.5,8,2,0.5,3)$,  these parameters are not fundamental physical constants derived analytically, 
but rather representative values chosen through empirical fitting. The analytical analysis only determines the envelope function form describing the influence of the measurement strength and the localized  potential strength on the entanglement growth rate, while the explicit numerical values of the parameters cannot be obtained in closed form. The selected set therefore serves as an effective choice that provides a reasonable description of the numerical phase boundaries and allows us to generate the plots.

\section{Free fermions with Anderson disorder under measurement}

We here complement the results for the SP and QPP by presenting the phase diagram of a one-dimensional chain of free fermions subject to Anderson-type disorder and continuous measurement. We consider spinless free fermions on a one-dimensional lattice with nearest-neighbor hopping $J$ in the presence of uncorrelated on-site disorder as shown in Eq.~\eqref{hami}, We set $J=1$ as the unit of energy and impose the same boundary conditions and filling as in the main text. In the Fig.~\ref{Dphase}, strong measurement or strong disorder yields an area-law phase with $\bar{c}_{\mathrm{eff}} \approx 0$, while weak measurement and weak disorder produce a logarithmic-entanglement phase with $\bar{c}_{\mathrm{eff}} > 0$.  
The phase boundary predicted by Eq.~\eqref{fs} in the main text is determined by the competition between the measurement-induced disentangling rate and the effective propagation scale set by localization, and it agrees very well with the numerically obtained transition line. Using the same parameters as for the quasiperiodic potential, we achieve an accurate fit to the phase boundary in the disordered phase diagram. This further supports our conclusion that the phase boundary depends only on the localization characteristics of the wave function; that is, as long as the lattice potential leads to the same degree of localization, the phase boundaries will be similar, largely independent of the specific form of the potential used in the model.
\section{Explanation of the measurement process}
\begin{figure}[bt]
\hspace*{-0.53\textwidth}
\includegraphics[width=0.5\textwidth]{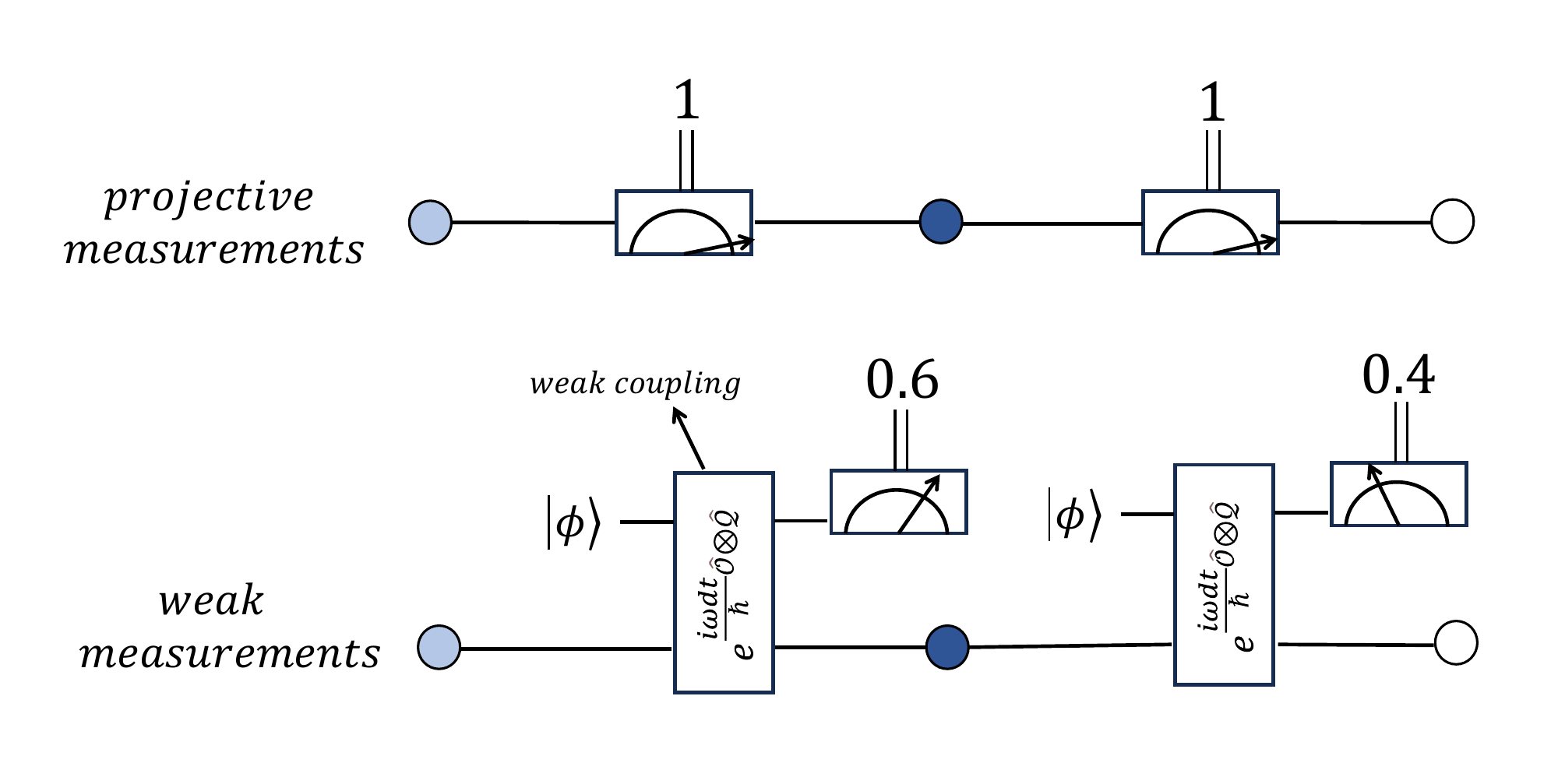}
\caption{ When a projective measurement is performed on an arbitrary quantum state, the system collapses into an eigenstate of the measurement operator. Consequently, repeated measurements of the same observable will necessarily yield identical outcomes. In contrast, a weak measurement is implemented by weakly coupling the system to an ancillary qubit $\phi$, which is then read out. This procedure extracts only partial information about the system without fully projecting it onto an eigenstate. Therefore, a subsequent measurement of the same observable may yield a different outcome and leave the system in a different post-measurement state. }
\label{figms}
\end{figure}

We are monitoring the local particle number operator $n_j = c_j^\dagger c_j$ through continuous weak measurement as shown in Fig.~\ref{figms} .
Specifically, different from projective measurement, the core idea of weak measurement is to employ a weak coupling between the system and the measurement apparatus, such that each individual measurement only slightly perturbs the system state while still extracting partial information. To formalize this process, we introduce a continuous measurement outcome $x$ and construct a series of Kraus operators:
\begin{equation}
M(x)=\frac{1}{\left(2 \pi \sigma^2\right)^{1 / 4}} e^{-\frac{(x-\omega \hat{O} d t)^2}{4 \sigma^2}} ,
\end{equation}
where, $\sigma$ describes the width of the probe initial wave function. These operators satisfy the continuous completeness relation $\int d x M^{\dagger}(x) M(x)=I$. For a single measurement on an initial state $|\psi\rangle$, the probability density of obtaining the result $x$ is given by $p(x)=\langle\psi| M^{\dagger}(x) M(x)|\psi\rangle$, and the normalized post-measurement state is updated to
\begin{equation}
|\psi(x)\rangle=\frac{M(x)|\psi\rangle}{\sqrt{p(x)}}.
\end{equation}
In Kraus representation, to track the evolution of a monitored quantum system at the pure state level and capture the randomness introduced by measurements, one often employs the quantum trajectory method, that is, the Stochastic schr¨odinger equation. The evolution of the system over a very short time interval $d t$ is represented by a set of Kraus operators $\left\{M_\alpha\right\}$, so that the density matrix evolves as:
\begin{equation}
\rho(t+d t)=\sum_\alpha M_\alpha(d t) \rho(t) M_\alpha^{\dagger}(d t).
\end{equation}
These operators can be divided into two categories:
\begin{equation}
M_0=I-i H_{\mathrm{eff}} d t, \quad M_k=\sqrt{\gamma d t} L_k.
\end{equation}
Here, $H_{\text {eff }}=H-\frac{i \gamma}{2} \sum_{k \geq 1} L_k^{\dagger} L_k$ is the non-Hermitian effective Hamiltonian and $L_k$ is the jump operator. $M_0$ and $M_k$ denote the no-jump operator (sometimes referred to as the non-Hermitian evolution operator) and the jump operator, respectively. In the discrete picture, during each small time step $d t$, the system either evolves without a jump or experiences a single jump. The probability for a jump is determined by a corresponding expression
\begin{equation}
p_{k \geq 1}=\langle\psi(t)| M_k^{\dagger} M_k|\psi(t)\rangle \approx \gamma d t\langle\psi(t)| L_k^{\dagger} L_k|\psi(t)\rangle .
\end{equation}
If a jump occurs, the state vector is updated according to the jump operator, i.e.,
\begin{equation}
|\psi(t+d t)\rangle=\frac{M_k|\psi(t)\rangle}{\sqrt{p_{k \geq 1}}},
\end{equation}
and if no jump occurs, the state vector evolves under the no-jump operator:
\begin{equation}
|\psi(t+d t)\rangle=\frac{M_0|\psi(t)\rangle}{\sqrt{p_0}},
\end{equation}
where $p_0=1-\sum_{k \geq 1} p_k$ is the probability for no jump. This framework effectively captures the stochastic nature of the evolution in a monitored quantum system.

In the continuous limit $d t \rightarrow 0$, the evolution process can be described by an Itô stochastic differential equation. One of the most common formulations is known as the quantum jump unraveling. In this approach, the state evolves continuously under a non-Hermitian effective Hamiltonian until a quantum jump occurs. The probability of a jump in an infinitesimal time interval is determined by the jump operator, and when a jump occurs, the state is updated according to that operator. Between jumps, the state follows a deterministic (but non-unitary) evolution. Formally, the quantum jump stochastic Schrödinger equation can be written as
\begin{align}
d|\psi(t)\rangle &= -i H_{\mathrm{eff}}|\psi(t)\rangle dt \nonumber \\
&\quad + \sum_{k \geq 1}\left(\frac{L_k|\psi(t)\rangle}{\| L_k|\psi(t)\rangle \|}-|\psi(t)\rangle\right) d N_k(t),
\label{se}
\end{align}
where $d N_k(t)$ is the set of Poisson stochastic increment, this is the Eq.~\eqref{eq3} and the Born rule is indeed encoded in the norm of the jump operators $\| L_k|\psi(t)\rangle \|$.
The Poisson increments $dN_k(t)$ stochastically determine whether a jump occurs in the interval $dt$ (in the simulations, we set $dt = 0.05$.
). The probability for a jump is given by
\begin{equation}
p_k = \langle \psi(t)| L_k^\dagger L_k |\psi(t)\rangle dt = \|L_k|\psi(t)\rangle\|^2 dt,
\end{equation}
which is precisely the Born probability. The no-jump probability is $p_0 = 1 - \sum_k p_k$. Thus, the randomness of each trajectory is fully governed by the Born rule, not by any artificial filtering. Physical observables are then obtained by averaging over the full ensemble of trajectories, rather than by keeping only a specific subset. 
%%%%%%%%%%%%%%%%%   BIB    %%%%%%%%%%%%%%%%%%%%%
\bibliography{bibliography}

\end{document}